\documentclass[twocolumn,showpacs,preprintnumbers,amsmath,amssymb]{revtex4}
\usepackage[dvips]{graphicx}
\usepackage{dcolumn}

\begin{document}

\title
{Vortex Lattice Melting Line in Superconductors with Paramagnetic Pair-Breaking}

\author{Dai Nakashima and Ryusuke Ikeda}

\affiliation{%
Department of Physics, Kyoto University, Kyoto 606-8502, Japan
}

\date{\today}

\begin{abstract} 
Recent experiments on the Iron-based superconductor FeSe in a high magnetic field have suggested the presence of {\it both} the fluctuation-induced vortex liquid regime and a Fulde-Ferrell-Larkin-Ovchinnikov (FFLO) vortex lattice. To get a general picture on the magnetic phase diagram in type II superconductors with strong superconducting (SC) fluctuation {\it and} strong paramagnetic pair-breaking (PPB) such as FeSe, the vortex lattice melting curve $H_m(T)$ is theoretically investigated in the situations where a FFLO state is expected to occur. In general, PPB tends to narrow the vortex liquid regime intervening between $H_{c2}(T)$ and $H_m(T)$. In particular, the vortex liquid regime is found to rapidly shrink upon entering, by cooling, the temperature range in which the FFLO state with a periodic modulation parallel to the magnetic field is stable in the mean field theory. Based on the present results, the high field SC phase diagrams of FeSe in the parallel and perpendicular field configurations are discussed. 
\end{abstract}
 


\maketitle

\section{Introduction}
In type II superconductors, two kinds of spatial modulations are created in the superconducting (SC) states by an applied magnetic field. One is the vortex structure \cite{Abrikosov} protected by the flux quantization, which is a topological condition in real space, and the other is a kind of Fulde-Ferrell-Larkin-Ovchinnikov (FFLO) spatial modulation \cite{FF,LO} which is usually supported by a field-induced splitting between the up-spin and down-spin portions of the Fermi surface. The former is a consequence of the orbital-pair breaking effect of the magnetic field, while the latter is that of the Pauli paramagnetic pair-breaking (PPB). 
 
The presence of the vortices {\it qualitatively} changes the nature of the SC fluctuation: The fluctuation changes the SC transition line (or, the upper critical field) $H_{c2}(T)$ in the mean field theory to a crossover line between the regions with strongly interacting SC fluctuations and with weakly interacting fluctuations \cite{IOT89,FFH}. The former region below the $H_{c2}(T)$ curve is often called as the vortex liquid regime. In clean limit, the true SC transition line in a magnetic field is the position of the vortex lattice melting in the field v.s. temperature ($H$-$T$) phase diagram and separates the vortex liquid regime from the vortex lattice or solid \cite{FFH,BG}. 
So far, any superconducting material, including the high $T_c$ cuprates, with a broad vortex liquid regime has not shown PPB-induced SC phenomena. On the other hand, most of searches for a FFLO state have been performed so far in situations where the effects of fluctuating vortices are 
invisible \cite{Bianchi,Kenzel}. 

Recent experiments on the Iron-based quasi two-dimensional (Q2D) superconductor FeSe have shown that, in both cases with a field parallel to the SC layers (${\bf H} \perp c$) \cite{Kasa20} and a field perpendicular to the layers (${\bf H} \parallel c$) \cite{Kasa21}, the fluctuation-induced vortex liquid regime coexists with a PPB-induced novel high field SC phase in the same phase diagram. To the best of our knowledge, this is the first material with a possible FFLO vortex phase existing just below a well-defined vortex liquid regime. One remarkable feature on the high field phase diagram of FeSe is that the mean field $H_{c2}$-transition in FeSe is apparently continuous in both field configurations \cite{Kasa20,Kasa21,Hardy} in contrast to the corresponding ones in CeCoIn$_5$ discussed repeatedly previously \cite{Bianchi,Kenzel,Kenzelmann}. On the other hand, the nature of the transition between the high field SC phase and the low field vortex solid differ between the two field configurations \cite{Kasa20,Kasa21}. 

In the present work, we investigate how the position of the vortex lattice melting of a Q2D type II superconductor is affected by strong PPB. Throughout the present work, we focus on the configuration with an applied magnetic field perpendicular to the basal plane of the superconductor \cite{AI03}, and the vortex lattice structure is assumed to be the conventional hexagonal one. The melting line will be examined in the following two manners. First, the elastic free energy of the vortex lattice is derived by taking account of the possibility of a formation of a FFLO spatial modulation parallel to the applied magnetic field, and the Lindemann criterion on the melting line \cite{BG,Moore} is derived based on the obtained elastic energy. Second, the melting curve will also be examined within the Ginzburg-Landau (GL) fluctuation analysis \cite{Hikami} by comparing the free energy obtained by approaching from the normal state with the free energy in the fluctuating vortex solid. In both of the two methods, the obtained vortex liquid region becomes narrower with increasing PPB and, in particular, upon entering the low temperature range in which the FFLO state is stable in the mean field approximation. Qualitatively, this result is consistent with the previous argument \cite{AI03} based on the Lindemann criterion that the melting line should merge with the $H_{c2}$-line in the temperature range where the mean field $H_{c2}$-transition is of first order. On the other hand, it will also be clarified that the amplitude fluctuation makes the validity of this argument vague. 

The present paper is organized as follows. The vortex lattice melting field $H_m(T)$ is derived according to the Lindemann criterion in sec.II and based on comparison between the obtained free energies in sec.III. In sec.IV, numerically obtained phase diagrams resulting from the methods in sec.II and III are compared with each other. In sec.V, our results are summarized, and their relevances to experimental phase diagrams of FeSe are discussed in details. In Appendix, the microscopic derivation of the GL model is reviewed. 

\section{Lindemann Criterion}

The Ginzburg-Landau (GL) Hamiltonian we use in the main text of this paper takes the form 
\begin{eqnarray}
{\cal H}_{\rm GL} &=& N(0) \int d^3{\bf r} \biggl[ \Delta^* (a_0 + B (-\partial_z^2) + C \partial_z^4 ) \Delta + \frac{V_4}{2} |\Delta|^4 \nonumber \\
&+& \frac{V_6}{3} |\Delta|^6 \biggr] 
\label{GL}
\end{eqnarray}
under a magnetic field parallel to the $z$-axis, 
where $N(0)$ is the electronic density of states per spin in the normal state. In eq.(\ref{GL}), it was assumed that the order parameter $\Delta$ is already in its lowest Landau level ($n=0$ LL) subspace, and that the mean field $H_{c2}(T)$ line is given by $a_0=0$ as far as $V_4$ is not negative. A microscopic derivation of eq.(\ref{GL}) in the presence of PPB was performed previously \cite{AI03,RI07}, and the details of the coefficients appearing in eq.(\ref{GL}) are given in Appendix. In low enough fields and at higher temperatures where PPB is negligible, the coefficients $B$ and $V_4$ are positive, while $C$ and $V_6$ are negative and are conventionally assumed to be zero in such cases. On the other hand, in the situations with strong PPB, the coefficients $C$ and $V_6$ change their sign with decreasing the temperature and with increasing the field, and, upon cooling further, $V_4$ and $B$ become negative in higher fields. A negative $B$ implies that a FFLO state with a modulation parallel to the field tends to form upon cooling. On the other hand, a negative $V_4$ implies that the mean field $H_{c2}$ transition is of first order, although, in real systems with SC fluctuation, this first order transition never occurs in reality and is reflected just as a crossover. As shown in Ref.\cite{AI03}, this crossover may be accompanied by a hysterisis signaling the mean field discontinuous transition. Below, it will be examined how these PPB effects affect the vortex lattice melting line $H_m(T)$. 

The mean field vortex lattice solution is obtained in a conventional manner \cite{Abrikosov}. First, as far as any SC fluctuation is absent, the non-Gaussian terms of eq.(\ref{GL}) can be rewritten in the manner 
\begin{equation}
\biggl\langle \frac{V_4}{2} |\Delta|^4 + \frac{V_6}{3} |\Delta|^6 
\biggr\rangle_s 
= \frac{{\tilde V}_4}{2} (\langle |\Delta|^2 \rangle_s)^2  + \frac{{\tilde V}_6}{3} (\langle |\Delta|^2 \rangle_s)^3 
\label{replace}
\end{equation}
where $\langle \,\,\,\, \rangle_s$ denotes the spatial average, $S$ is the system area in the plane perpendicular to the field, ${\tilde V}_4=\beta_{\rm A} V_4$, and ${\tilde V}_6=\gamma_{\rm A} V_6$ with 
\begin{eqnarray}
\beta_{\rm A} &=& \frac{\langle |\Delta|^4 \rangle_s}{(\langle |\Delta|^2 \rangle_s)^2} = 1.1596, \nonumber \\
\gamma_{\rm A} &=& \frac{\langle |\Delta|^6 \rangle_s}{(\langle |\Delta|^2 \rangle_s)^3} = 1.4230.
\end{eqnarray}
Hereafter, as a possible PPB-induced spatial modulation of $\Delta$ developing along the applied field direction, the helical phase modulation will also be included. Then, the mean field solution with the in-plane triangular vortex lattice structure and the out-of-plane helical phase modulation is given by $\Delta_0=\alpha_0 \varphi({\bf r}_\perp|0) e^{iq_m z}$ with $q_m^2 = \theta(-B) |B|/2C$, where $\varphi({\bf r}_\perp|0)$ is the Abrikosov solution \cite{Abrikosov,Moore,Eilenberger,RI591740}
\begin{equation}
\varphi({\bf r}_\perp|0) = \sqrt{\frac{k r_H}{\pi^{1/2}}} \sum_{n=-\infty}^\infty \exp\biggl[-\frac{y^2}{2 r_H^2} + {\rm i}kn \biggl(x+\frac{\pi}{2k}n -iy \biggr) \biggr]
\label{magBloch}
\end{equation} 
constructed in the $n=0$ LL and in the Landau gauge ${\bf A}=-Hy{\hat x}$ and satisfying the normalization condition $\langle |\varphi({\bf r}|0)|^2 \rangle_s=1$, $k=\pi^{1/2} 3^{1/4} r_H^{-1}$, $r_H = \sqrt{\phi_0/(2 \pi H)}$, $\phi_0= \pi \hbar/|e|$ is the flux quantum, and ${\bf r}_\perp =$ ($x$, $y$) denotes the 2D coordinate. The value of $\alpha_0^2$ minimizing the free energy, eqs.(\ref{GL}) and (\ref{replace}), 
is given by 
\begin{equation}
\alpha_0^2=\frac{|{\tilde V}_4|}{2 {\tilde V}_6} \biggl( - s_4 + \sqrt{ 1 - {\overline a}'_0} \biggr), 
\label{MFamp}
\end{equation}
where 
\begin{equation}
{\overline a}'_0=\frac{4 {\tilde V}_6}{{\tilde V}_4^2} a'_0 = \frac{4 {\tilde V}_6}{{\tilde V}_4^2} \, (a_0 - C q_m^4), 
\label{abardash}
\end{equation}
and $s_4 = V_4/|V_4|$. 
The mean field SC transition at $H_{c2}$ is of second order when $s_4 > 0$, while it is of first order when $s_4 < 0$. The resulting $H_{c2}(T)$-line is given 
by $a'_0=0$ for the former, while it is given by ${\overline a}'_0 = 3/4$ for the latter. Then, the free energy density of the mean field solution becomes 
\begin{equation}
f_{\rm MF} = - N(0) \frac{|{\tilde V}_4|^3}{12 {\tilde V}^2_6} \biggl[ \frac{1}{2} s_4 {\overline a}'_0 + ( 1 - {\overline a}'_0)^{3/2} 
- s_4(1 - {\overline a}'_0) \biggr].
\label{fMF}
\end{equation} 

Next, the elastic energy of the vortex lattice will be considered. As far as we restrict ourselves to the type II limit with no gauge field fluctuation incorporated, the elastic energy of the vortex lattice is obtained as the energy of the massless harmonic fluctuation within the $n=0$ LL around the vortex lattice solution (\ref{magBloch}) \cite{Moore,Eilenberger,RI591740}. Its derivation in the presence of PPB is sketched in Appendix. The resulting Hamiltonian of the massless mode becomes  
\begin{equation}
\delta {\cal H}_{\rm ph} = \frac{1}{2} \sum_{q, {\bf k}_\perp} (\rho_s q^2 + \sigma_s q^4 + C_{66} {\overline k}_\perp^4) |\delta \chi(q,{\bf k}_\perp)|^2, 
\label{phaseenergy}
\end{equation}
where $q$ is the wave number measured from $q_m$ in the $z$-direction \cite{AI03}, 
\begin{eqnarray}
\rho_s &=& 2 N(0) \, \alpha_0^2 \, |B| \, ( 1 + \theta(-B)), \nonumber \\
\sigma_s &=& 2N(0) \, C \, \alpha_0^2 \, \theta(C), 
\label{rhos}
\end{eqnarray}
and 
\begin{equation}
C_{66} = 2 N(0) \, \alpha_0^4 \, ( 0.119 \, V_4 + 0.276 \, \alpha_0^2 \, V_6). 
\label{shearmod}
\end{equation}
The first term of eq.(\ref{shearmod}) coincides with the result in the previous work \cite{Moore}. 
In fact, in low enough fields ($H \ll H_{c2}(0)$) where PPB is negligible so that $B$ approaches $\xi_0^2$, we have $\rho_s r_H^{-4} \simeq H^2/(4 \pi \lambda^2(T))$, and eq.(\ref{phaseenergy}) becomes the energy of the shear elastic fluctuation in type II limit by assuming, as mentioned earlier, $\sigma_s$ and $V_{6}$ to be zero: 
\begin{equation}
\delta {\cal H}_{\rm ph} \simeq \frac{1}{2} \sum_{q,{\bf k}_\perp} \biggl( \frac{H^2}{4 \pi \lambda^2(T) k_\perp^2} q^2 + C_{66} k_\perp^2 \biggr)|s_{\rm T}(q, {\bf k}_\perp)|^2 
\label{lowHshearenergy}
\end{equation}
where $r_H^2 (\nabla \times {\hat z}) \delta \chi$ was identified with the transverse component ${\bf s}_{\rm T}$ of the vortex displacement field \cite{Moore,RI591740}. Here, $\xi_0$ and $\lambda(0)$ are, respectively, the zero temperature coherence length and penetration depth defined within the GL theory in low fields. 

Based on eq.(\ref{shearmod}), one might be afraid of whether $C_{66}$ approaches zero upon cooling in the region where $V_4$ is negative. However, one can check that $C_{66}$ remains positive even upon approaching the first order $H_{c2}$-transition line from below.  

The Lindemann criterion for determining the vortex lattice melting line $H_m(T)$ becomes \begin{equation}
\langle s_{\rm T}^2 \rangle = \int_q \int_{{\bf k}_\perp} \frac{T k_\perp^2 r_H^4}{\rho_s q^2 + \sigma_s q^4 + C_{66}k_\perp^4 r_H^4} = c_{\rm L}^2 r_H^2, 
\label{Lindemann} 
\end{equation}
where the ${\bf k}_\perp$-integral is performed by setting the first Brillouin Zone of the vortex lattice to be, for simplicity, circular. The constant parameter $c_{\rm L}$ needs to be determined phenomenologically or empirically. The $q$-integral can be performed analytically, and the Lindemann criterion in the presence of PPB becomes 
\begin{widetext}
\begin{equation}
\frac{T}{\sqrt{\rho_s C_{66}}}  = c_{\rm L}^2 \frac{\phi_0}{H} \biggl[ 1 + \sqrt{1 + 4\sqrt{\frac{C_{66} \sigma_s}{\rho_s^2}}} \biggr]. 
\label{Lindemannresult}
\end{equation}
\end{widetext}
Based on the above-mentioned fact that, when the $H_{c2}$-transition is of first order, $C_{66}$ does not vanish on approaching the $H_{c2}$-line from below, it is clear that the equality in eq.(\ref{Lindemannresult}) is not satisfied at low enough $T$. It inevitably leads to the argument that, in the temperature range where the mean field SC transition at $H_{c2}$ is of first order, the melting curve $H_m(T)$ tends to merge with the $H_{c2}$-curve at a finite temperature, and consequently that the vortex liquid regime in a superconducting material with strong PPB tends to disappear at the temperature on cooling \cite{AI03,Hardy}. The validity of this argument based on the elastic theory will be discussed again in sec.IV. 

\section{Approach based on free energy evaluation}

It is useful to compare the result on the melting line $H_m(T)$ in sec.II with that following from a different approach in order to see to what extent the Lindemann criterion in sec. II is reliable. In this section, we try to obtain $H_m(T)$ based directly on calculating the fluctuation free energy \cite{Hikami}. 

Imagine that one starts from the normal phase. The fluctuation propagator ${\cal D}_{p,q} = \langle |\Delta_{0, p,q}|^2 \rangle$ of the $n=0$ LL modes $\Delta_{0,p,q}$ in the Gaussian approximation is defined as 
\begin{equation}
{\cal D}_{p,q} = \frac{T}{N(0)(a_0 + B q^2 + C q^4)}. 
\label{Gaussprop}
\end{equation}
in the case of the GL Hamiltonian in eq.(\ref{GL}). Here, as in sec.II, the Landau gauge was chosen for the vector potential. Due to the degeneracy in each LL, the r.h.s. of eq.(\ref{Gaussprop}) is independent of the wavenumber $p$ defined in the plane perpendicular to the field. 

To extend this fluctuation propagator to the case with the mode-coupling terms, the renormalized mass $\mu$ will be introduced as 
\begin{equation}
\mu = a_0 + \Sigma. 
\label{Dyson}
\end{equation}
Then, eq.(\ref{Gaussprop}) is replaced by 
\begin{equation}
{\cal D}_{p,q} = \frac{T}{N(0)(\mu + B q^2 + C q^4)}. 
\label{prop}
\end{equation}

In general, the self energy $\Sigma$ is constructed based on the perturbation expansion with respect to the mode-coupling terms, and a non-Gaussian or renormalized theory of the SC fluctuation is formulated \cite{IOT89,Hikami}. In contrast to the case \cite{Hikami,Ruggeri} studied thoroughly so far with no PPB, however, it is extremely cumbersome to study the large order behavior of the perturbation series in the present case with PPB where the sixth order term and the quartic gradient term need to be taken into account in the starting GL free energy. Fortunately, the feature deep in the vortex liquid regime that the {\it amplitude} $|\Delta|$ of the SC order parameter is well defined holds in the present case with strong PPB. This feature is well approximated based on the self-consistent Hartree-Fock approximation \cite{IOT90,AI17} in which $\mu$ is independent of $q$. Below, the Hartree-Fock approximation will be used by expecting the parameter dependences of the resulting phase diagram to be comparable with those of the result of the Lindemann criterion. 
Then, the renormalized mass $\mu$ is determined by the self-consistent equation (\ref{Dyson}) with 
\begin{eqnarray}
\Sigma &=& V_4 \frac{H}{N(0) \phi_0} \frac{T}{\sqrt{\mu(B+2\sqrt{C \mu})}} + \frac{3 V_6}{2} \biggl(\frac{H}{N(0) \phi_0} \biggr)^2 \nonumber \\
&\times& \frac{T^2}{\mu(B+2\sqrt{C \mu})}. 
\label{selfenergy}
\end{eqnarray}
We note that, in the case with a negative $B$, i.e., in the temperature range where the ordered phase in lower fields is a FF vortex lattice, not $\mu$ but 
\begin{equation}
\mu' \equiv \mu - \frac{B^2}{4 C}
\label{rmubelow}
\end{equation}
plays the role of the renormalized mass, reflecting the fact that the corresponding bare mass is not $a_0$ but 
$a'_0$ defined in eq.(\ref{abardash}). 

Next, to write down the expression of the fluctuation free energy density $f_>$ in the normal phase, the exact expression on the mean squared average of the SC order parameter $\langle |\Delta|^2 \rangle$ 
\begin{equation}
N(0) \langle |\Delta|^2 \rangle = \frac{\partial f_>(a_0)}{\partial a_0} 
\label{msqav}
\end{equation}
will be used. Then, in the Hartree-Fock approximation, the free energy density in $B > 0$ case becomes  
\begin{eqnarray}
f_> &=& \int_\infty^{a_0} d\varepsilon \frac{\partial f_>(\varepsilon)}{\partial \varepsilon} + f_0 \nonumber \\
&=& N(0) \int_{0}^\mu d\mu \, \langle |\Delta|^2 \rangle - \int_\infty^\mu d\mu \frac{\partial \Sigma}{\partial \mu} \frac{\partial f_>(a_0)}{\partial a_0} \nonumber \\
&=& \frac{HT}{\phi_0} \biggl[ \frac{\sqrt{B+2\sqrt{C \mu}} - \sqrt{|B|}}{\sqrt{C}}  - \frac{V_4}{4} \biggl(\frac{HT}{N(0) \phi_0} \biggr) \nonumber \\
&\times& \frac{1}{\mu(B+2\sqrt{C \mu})} - \frac{V_6}{2} \biggl(\frac{HT}{N(0) \phi_0} \biggr)^2 \nonumber \\
&\times& \frac{1}{[\mu(B + 2\sqrt{C \mu})]^{3/2}} \biggr]. 
\label{flucf}
\end{eqnarray}
In the above analysis, the expression 
\begin{equation}
N(0) \int_0^\infty d\mu  \langle |\Delta|^2 \rangle
\end{equation}
has been chosen as the constant $f_0$ in the first line of eq.(\ref{flucf}). Then, in $C \to 0$ limit, eq.(\ref{flucf}) is precisely the same as the corresponding expression in Ref.\cite{Hikami}. 

The corresponding expression to eq.(\ref{flucf}) in $B < 0$ case is obtained in almost the same manner, although $N(0) \int_0^\infty d\mu' \langle |\Delta|^2 \rangle$ is chosen as $f_0$ of the first line in eq.(\ref{flucf}). Then, the free energy density in $B<0$ case is given simply by adding $HT (|B|/C)^{1/2}/\phi_0$ to eq.(\ref{flucf}), that is,  
\begin{eqnarray}
f_> &=& \frac{HT}{\phi_0} \biggl[ \sqrt{\frac{4 \mu'}{2\sqrt{C \mu}+|B|}}  - \frac{V_4}{16} \biggl(\frac{HT}{N(0) \phi_0} \biggr) \frac{|B|+2\sqrt{C \mu}}{C \mu \mu'} \nonumber \\
&-& \frac{V_6}{16} \biggl(\frac{HT}{N(0) \phi_0} \biggr)^2 \biggl[\frac{|B|+2\sqrt{C \mu}}{C \mu \mu'} \biggr]^{3/2} \, \biggr]. 
\label{flucfneg}
\end{eqnarray}

In turn, the free energy density of the vortex lattice state will be examined \cite{Hikami}. To do this, the free energy arising from the harmonic excitations around the mean field vortex lattice will be added to $f_{\rm MF}$ given in eq.(\ref{fMF}). However, it is a well established fact within the approach based on the critical SC fluctuation that the shear elastic mode of the vortex lattice, i.e., the Goldstone mode, is a smaller correction compared with that of the amplitude fluctuation. Therefore, the free energy term resulting from eq.(\ref{phaseenergy}) can be neglected below. 
The Hamiltonian $\delta {\cal H}_{\rm amp}$ on the Gaussian amplitude fluctuation is presented in Appendix. In the same way as the derivation of the first term of $f_>$ in sec.II, the fluctuation correction $\delta f$ to $f_{\rm MF}$ resulting from $\delta {\cal H}_{\rm amp}$ is obtained in the form 
\begin{equation}
\delta f = \frac{HT}{\sqrt{2 C} \phi_0} \biggl( \, \sqrt{B+2 \sqrt{r_< C}} - \sqrt{B} \, \biggr), 
\end{equation}
for $B > 0$, and 
\begin{equation}
\delta f = \frac{HT}{\sqrt{2 C} \phi_0} \sqrt{2|B|+2 \sqrt{r_< C}} 
\end{equation}
for $B < 0$, where 
\begin{equation}
r_< \simeq \frac{{\tilde V}_4^2}{{\tilde V}_6} \biggl( 1 - {\overline a}'_0 - s_4 \sqrt{1 - {\overline a}'_0} \biggr). 
\end{equation}
In this manner, the free energy density based on the approach from lower temperatures is given by $f_< = f_{\rm MF} + \delta f$. 

After all, the melting line $H_m(T)$ is determined by the equality $f_>=f_<$ within the approach in this section. 

\section{Results on Magnetic Phase Diagram} 

\begin{figure}[t]
\begin{center}
{
\includegraphics[scale=0.45]{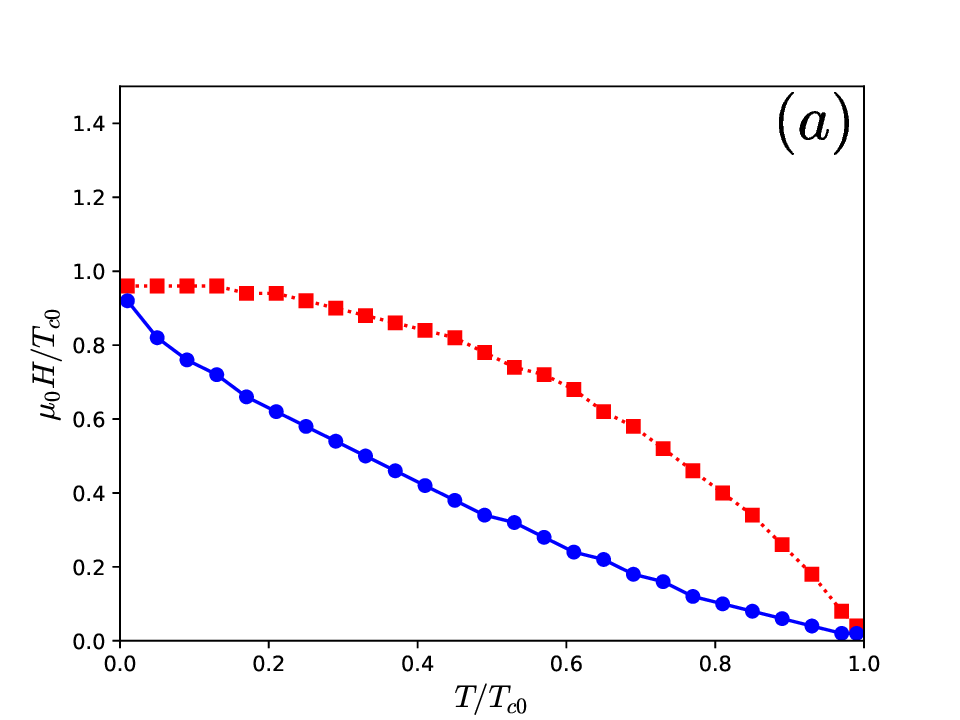}
\includegraphics[scale=0.45]{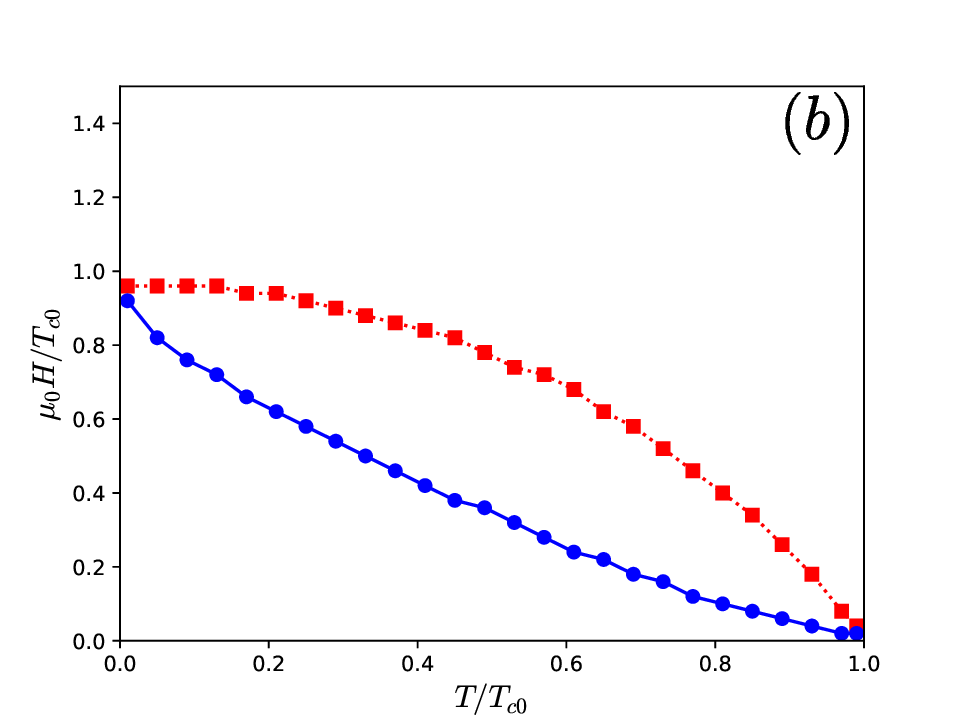}
}
\caption{(Color online) 
Field v.s. temperature ($H$-$T$) phase diagrams obtained (a) from eq.(\ref{Lindemannresult}) and (b) from the approach in sec.III. The used parameter values are $\alpha_{\rm M}=0.25$ and ${\tilde N}_0=0.1$. The red dotted curve with red square symbols denotes the crossover line corresponding to the mean field second order SC transition line $H_{c2}(T)$, while the blue solid curve denotes the vortex lattice melting line which is the genuine SC transition line in clean limit. } 
\label{fig.1}
\end{center}
\end{figure}

In this section, the melting transition lines of the $n=0$ LL vortex lattice following from the two methods explained in the preceding sections will be compared with each other to obtain a generic picture on the 
magnetic ($H$-$T$) phase diagram of the type II superconductors with moderately strong PPB. To make understanding of the parameter dependences of the phase diagram easier, it will be useful to rewrite the formula eq.(\ref{Lindemannresult}) based on the Lindemann criterion 
in terms of dimensionless variables in the form 
\begin{equation}
\frac{T}{4 \pi T_{c0}}\frac{\xi_0^2}{r_H^2} = {\tilde N}_0 \biggl(\frac{c_{\rm L} \alpha_0}{T_{c0}} \biggr)^2 g \, \biggl[ |b|^{1/2} + \sqrt{|b| + 4|c|^{1/2} g} \biggr], 
\label{scaledLindemann}
\end{equation}
where 
\begin{equation}
g = \sqrt{0.119 \alpha_0^2 {\tilde V}_4 + 0.276 \alpha_0^4 {\tilde V}_6}, 
\end{equation}
and 
\begin{equation}
{\tilde N}_0 = N(0) T_{c0} \xi_0^2 \xi_{0, \parallel} = \frac{0.003}{ \sqrt{\rm Gi} }
\end{equation}
expressed in terms of the Ginzburg number 
\begin{eqnarray}
{\rm Gi} &=& 2 \biggl(\frac{7 \zeta(3)}{64 \pi^3 N(0) T_{c0} \xi_0^2 \xi_{0, \parallel}} \biggr)^2 \nonumber \\
&=& 2 \biggl(\frac{7 \zeta(3)}{32 \pi^2} \, \frac{(\lambda(0))^2}{\Lambda(T_{c0}) \xi_{0, \parallel}} \biggr)^2 
\, \frac{|\Delta(0)|^4}{T_{c0}^4} 
\end{eqnarray}
measuring the strength of the thermal fluctuation is the scaled DOS, $\xi_0$ and $\xi_{0,\parallel}$ are the in-plane and out-of-plane coherence lengths of a Q2D material, $|\Delta(0)|$ is the zero temperature energy gap, $\Lambda(T)=\phi_0^2/(16 \pi^2 T)$ is the thermal length \cite{FFH}, and the dimensionless coefficients, $b=B/\xi_{0, \parallel}^2$ and $c=C/\xi_{0, \parallel}^4$, are given in (\ref{bc}) in Appendix. 
Based on eq.(\ref{scaledLindemann}), $H_m(T)$ will be discussed hereafter as a function of the fluctuation strength $1/{\tilde N}_0$ and the PPB strength, i.e., the Maki parameter \cite{Maki} 
\begin{equation}
\alpha_{\rm M}= \frac{\mu_0 H_{c2}^{({\rm orb})}(T=0)}{2 \pi T_{c0}}
\end{equation}
which is incorporated in the dimensionless GL coefficients $a_0$, $b$, $c$, $\alpha_0/T_{c0}$, $T_{c0}^2 V_4$, and $T_{c0}^4 V_6$, where $\mu_0 H$ is the Zeeman energy for a single quasiparticle, and $H_{c2}^{({\rm orb})}(T)$ is the $H_{c2}(T)$-line in ${\bf H} \parallel c$ in the absence of PPB. 
Similarly, the contributions to the free energy density introduced in sec.III, $f_{\rm MF}$, $f_>$, and $\delta f$, are also described in terms of the dimensionless GL coefficients parameterized by ${\tilde N}_0$ and $\alpha_{\rm M}$. 
In all of the $H$-$T$ phase diagrams to be discussed below, the temperature $T$ and the field strength $H$ are expressed in the units of $T_{c0}$ and the Pauli-limiting field $T_{c0}/\mu_0 = H_{c2}^{({\rm orb})}(T=0)/(2 \pi \alpha_{\rm M})$, respectively. 

Hereafter, the parameter values $\alpha_{\rm M}=0.25$ and $0.75$ will be used together with ${\tilde N}_0 = 0.1$ and $1.0$. Corresponding to the Gi-value of FeSe noted in Ref.\cite{Koshelev}, we have 
\begin{equation}
{\tilde N}_0=0.12.
\label{n0}
\end{equation}

\begin{figure}[t]
\begin{center}
{
\includegraphics[scale=0.45]{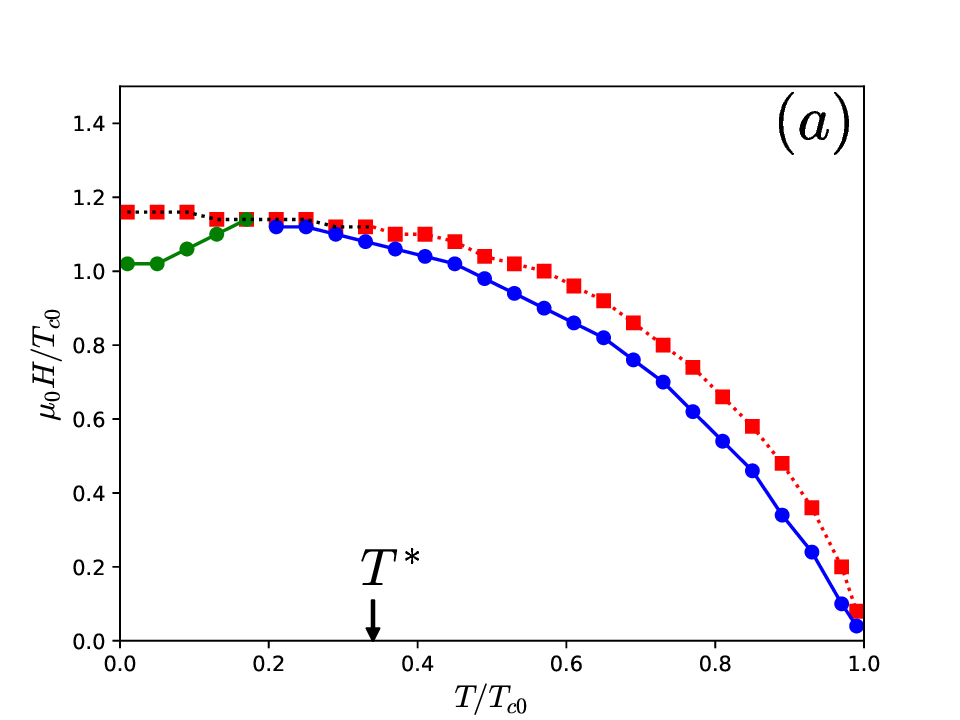}
\includegraphics[scale=0.45]{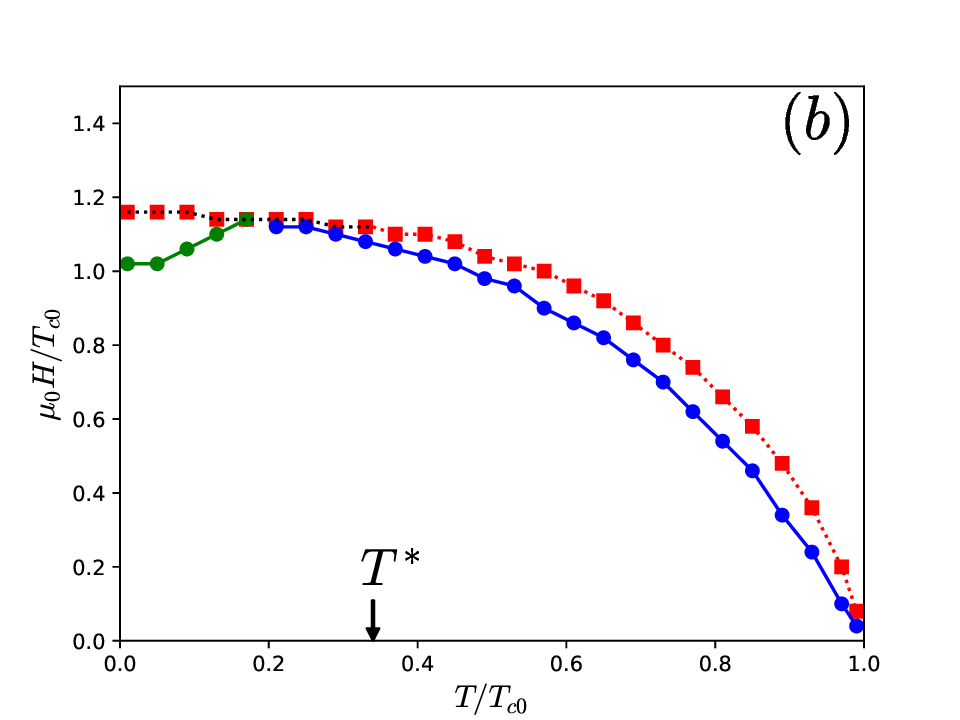}
}
\caption{(Color online) Field v.s. temperature ($H$-$T$) phase diagrams obtained (a) from eq.(\ref{Lindemannresult}) and (b) from the approach in sec.III. The used parameter values are $\alpha_{\rm M}=0.75$ and ${\tilde N}_0=1.0$. The $H_{c2}(T)$-curve (red square symbols) consists of the mean field second order transition line (red dotted curve) at which ${\overline a}'_0=0$ above $T^*$ (i.e., when $V_4 > 0$) and the mean field first order transition line (the black-dotted one) at which ${\overline a}'_0=3/4$ below $T^*$. The green solid curve at which the coefficient $B$ changes its sign is the transition line within $n=0$ LL between the conventional Abrikosov vortex lattice and the FFLO vortex lattice at lower temperatures. The melting line is expressed by the blue solid curve. 
} 
\label{fig.2}
\end{center}
\end{figure}

Figures 1 and 2 are different examples of comparison between the two melting lines obtained from the methods introduced in the preceding two sections under a fixed set of ${\tilde N}_0$ and $\alpha_{\rm M}$ values. In both of Figs.1 and 2, the melting curves obtained in terms of the two methods nearly coincide with each other when using the Lindemann constant $c_{\rm L}=0.414$, suggesting that the parameter dependences determining $H_m(T)$ are similar to each other between the two methods. Hereafter, the value $c_{\rm L}=0.414$ will be commonly used in obtaining results based on the Lindemann criterion. It will be seen later that such agreement is also seen in the case with stronger PPB {\it and} stronger fluctuation at least outside the FFLO temperature range in which a FFLO state is predicted in the mean field theory to occur (see Fig.3 below). 

In Fig.1, a strong fluctuation strength ${\tilde N}_0^{-1} = 10$ and a relatively weaker PPB strength $\alpha_{\rm M} = 0.25$ are used, and, as in the case with no PPB, $H_m(T)$ is concave, i. e., a curve with a positive curvature in the $H$-$T$ phase diagram and approximately obeys the $n=0$ LL scaling \cite{IOT89,Moore,Hikami} at least in lower fields where $B >0$ and $C=0$. At low enough temperatures where $B$ ($> 0$) is small, and $C > 0$, the $n=0$ LL scaling is not satisfied any longer. Nevertheless, the deviation from the $n=0$ LL scaling seems to be unexpectedly small. In contrast, in Fig.2 with ${\tilde N}_0^{-1}=1.0$ and $\alpha_{\rm M}=0.75$, the vortex liquid regime is narrow, and $H_m(T)$ begins to follow the functional form of the $H_{c2}(T)$ curve. That is, $H_m(T)$ is convex and has a negative curvature, except in the close vicinity of $T_{c0}$, 
in the $H$-$T$ phase diagram. 

Note that the $H_m(T)$-curve in Fig.2 apparently merges with the $H_{c2}(T)$-curve at a higher temperature than the FFLO temperature range. This tendency implying a shrinkage of the vortex liquid regime at a finite temperature is in agreement with the argument noted at the end of sec.II \cite{AI03}. 
As is seen in Fig.3, however, this tendency becomes unclear when the fluctuation is stronger. 

\begin{figure}[t]
\begin{center}
{
\includegraphics[scale=0.45]{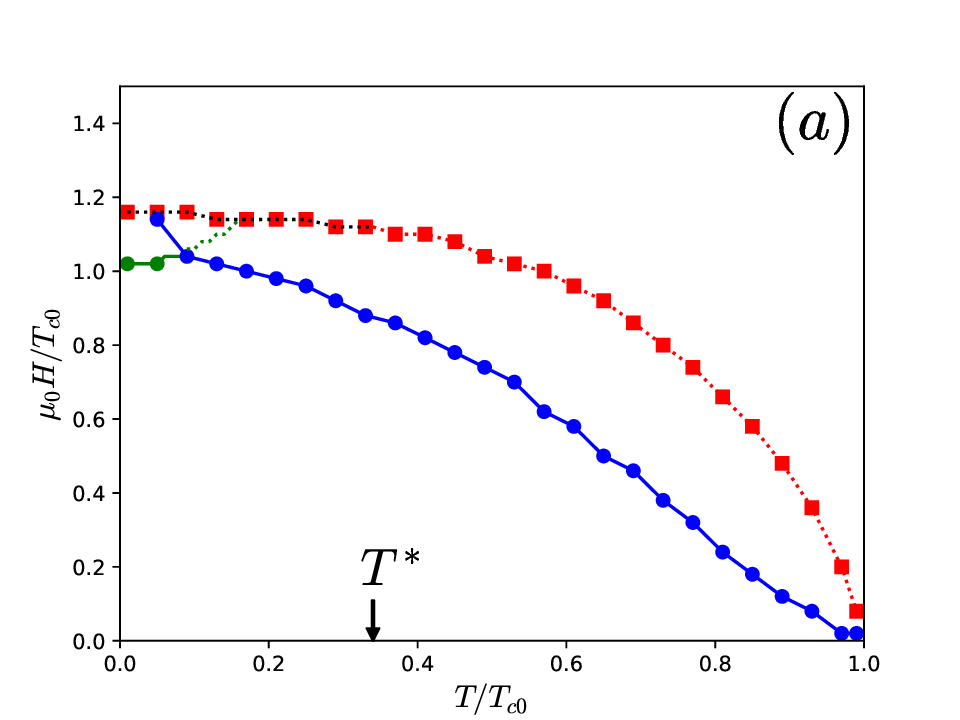}
\includegraphics[scale=0.45]{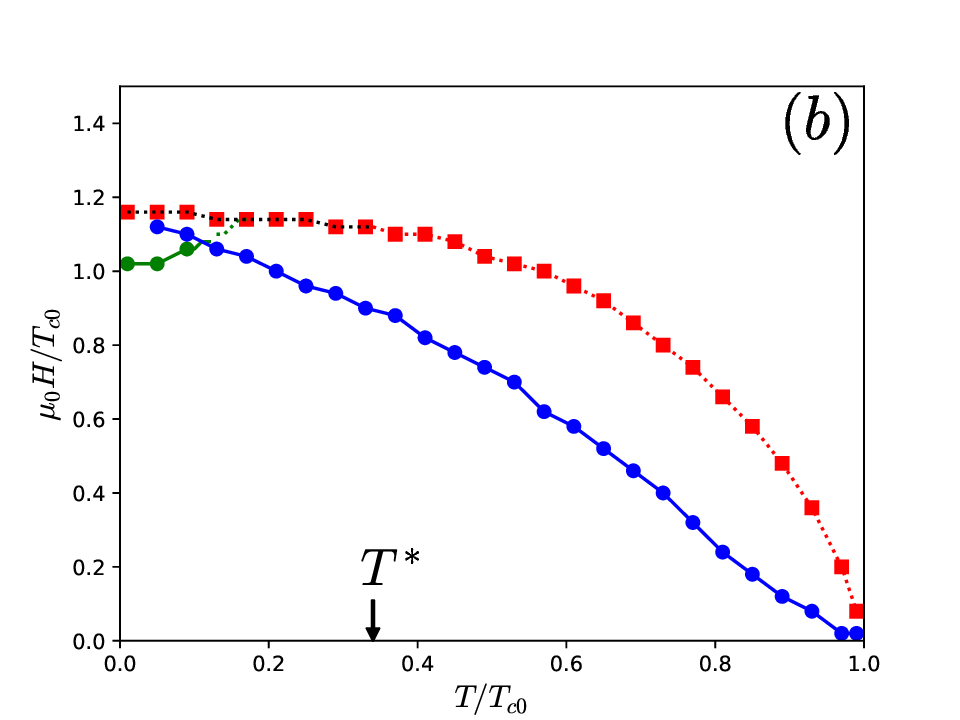}
}
\caption{(Color online) Field v.s. temperature ($H$-$T$) phase diagrams obtained (a) from eq.(\ref{Lindemannresult}) and (b) from the approach in sec.III. The content of the figures are the same as that of Fig.2 (a) and (b) except the use of the parameter values $\alpha_{\rm M}=0.75$ and ${\tilde N}_0=0.1$ in these figures. The portion (green dotted curve) of the mean field FFLO transition line in the vortex liquid regime is just a crossover line \cite{AI03}. 
} 
\label{fig.3}
\end{center}
\end{figure}
To understand the ${\tilde N}_0$ and $\alpha_{\rm M}$ dependences of the phase diagram in more details, the results on the phase diagram following from the two methods are shown in Fig.3, where the values ${\tilde N}_0^{-1}=10$ and $\alpha_{\rm M}=0.75$ are used. These two figures show typical phase diagrams in the case where both the fluctuation and PPB are moderately strong. One main feature is that, in spite of the strong fluctuation, the $H_m(T)$-curve is convex over most of the field range in the $H$-$T$ phase diagram. On the other hand, concave portions of $H_m(T)$ are seen close to $T_{c0}$ and in the FFLO temperature range. The former occurring in low enough fields where PPB weakly contributes is a consequence of the $n=0$ LL scaling \cite{IOT89,Moore,Hikami} $T - T_{c0} \sim (T H)^{2/3}$, while the latter reflects the shrinkage of the vortex liquid due to PPB \cite{AI03}. 

By comparing Figs.2 and 3 with each other, it is easily found that, at a fixed $\alpha_{\rm M}$, an increase of ${\tilde N}_0$ leads to a shrinkage of the vortex liquid regime while the convex $H_m(T)$-curve is kept. This feature suggests that, in superconductors with weak fluctuation, it is not easy to distinguish $H_m(T)$ from $H_{c2}(T)$ through experimental data. 

On the other hand, by comparing Figs.1 and 3 with each other, it is found that an increase of PPB makes the vortex liquid regime narrower. In particular, as mentioned above, it is commonly seen that the $H_m(T)$ curve becomes convex, like $H_{c2}(T)$, in the $H$-$T$ phase diagram, reflecting an enhanced role of PPB. Since ${\tilde N}_0^{-1}$ measures the strength of the fluctuation in zero field, the feature mentioned above implies that, in systems with moderately strong PPB, the actual fluctuation strength tends to be underestimated through experimental data in finite fields by, for instances, identifying the irreversibility line on which the resistivity vanishes with the $H_{c2}(T)$-line, because the melting curve lies quite close to the irreversibility line in most cases. 

Next, let us discuss the fate of the melting line in the FFLO temperature range. In Fig.3 (a) obtained based on the Lindemann criterion, the $H_m(T)$-line suddenly begins to approach the $H_{c2}(T)$-line on entering the FFLO temperature range. The origin of this sharp change seems to consist in the change in the coefficient of the $q^2$ term upon entering the FFLO temperature range by cooling (see eq.(\ref{rhos})). 
As seen in Fig.3 (b) obtained based on the free energy approach explained in sec.III, however, this change in $H_m(T)$ upon entering the FFLO range seems to become unclear as the amplitude fluctuation is incorporated. Therefore, the fate of $H_m(T)$ in the FFLO temperature range is sensitive to the details of its derivation, and it is not sufficiently understood at present whether $H_m(T)$ truly merges with $H_{c2}(T)$ in systems with strong enough fluctuation. 

\section{Summary and Discussion} 

In the present work, the vortex lattice melting curve in the type II superconductor with moderately strong paramagnetic pair-breaking (PPB) has been theoretically examined by assuming the vortex lattice to have the familiar hexagonal symmetry and hence to be described by the lowest ($n=0$) LL modes of the SC order parameter. The present result extends the previous works \cite{Moore,Houghton} constructing the Lindemann criterion of the vortex lattice melting to the cases with PPB. 

Below, let us discuss the magnetic phase diagrams of FeSe \cite{Kasa20,Kasa21,Hardy} as an example of application of the results in the preceding sections. 
The field configuration assumed in the present work corresponds to FeSe in ${\bf H} \parallel c$ where a nearly linear $H_m(T)$ in the temperature \cite{Hardy} and a high field SC (HFSC) phase \cite{Kasa21} were found. The nearly linear $H_m(T)$-curve is easily understood based on our Fig.1 and Fig.3. Clearly, the deviation from the concave melting curve suggesting the $n=0$ LL scaling \cite{Moore} is due to moderately strong PPB, and the region in which the $n=0$ LL scaling is correctly seen is limited to the low field range close to $T_{c0}$. It has been argued elsewhere \cite{AI19} that the low field behavior of the melting line is also affected by the strong-coupling effect due to the SC fluctuation itself in a system close to the so-called BCS-BEC crossover regime. 

\begin{figure}[b]
\begin{center}
{\includegraphics[scale=0.55]{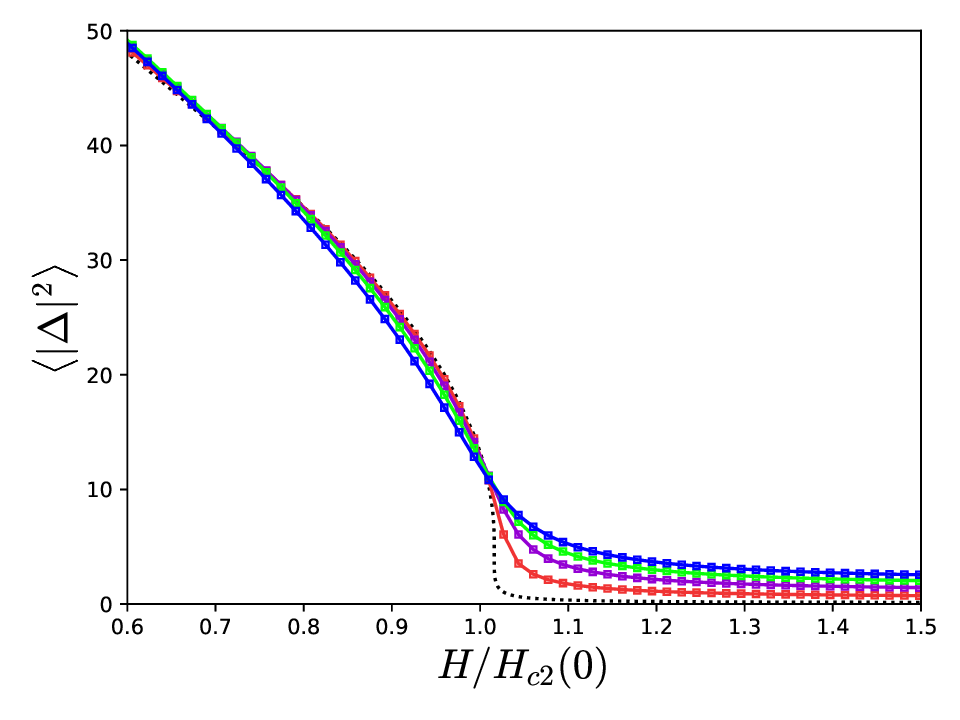}}
\caption{(Color online) Field dependences of the mean-squared amplitude $\langle |\Delta|^2 \rangle$ of the SC order parameter at the temperatures $T/T_{c0}=0.2$ (blue), $0.15$ (green), $0.10$ (purple), $0.05$ (red), and $0.01$ (black dotted curve) following from eqs.(\ref{Dyson}) and (\ref{selfenergy}) in terms of $\alpha_{\rm M}=0.8$ and ${\tilde N}_0=0.12$. Reflecting the fact that, when $\alpha_{\rm M}=0.8$, the $H_{c2}(T)$-line is the first order transition line in the mean field approximation in $t < 0.35$, $\langle |\Delta|^2 \rangle$ at $t=0.01$ where the thermal fluctuation is extremely weak shows a nearly discontinuous change at $H_{c2}(0)$. Inclusion of the quantum fluctuation neglected here would broaden even the $t=0.01$ curve. 
} 
\label{fig.4}
\end{center}
\end{figure}

On the other hand, extensive discussion is needed to understand the identity of the HFSC phase in ${\bf H} \parallel c$ \cite{Kasa21}. By taking account of the fact that, in ${\bf H} \parallel c$ where the fluctuation effect may become relatively strong, PPB does not become extremely strong compared with the orbital pair-breaking creating the vortices, we have at most two candidates of such a high field SC phase induced by PPB. One is the second lowest ($n=1$) LL vortex lattice \cite{Klein,Allan,Matsu}. Within the analysis based on the weak-coupling BCS approach for a {\it single-band} electronic model, the low temperature phase induced by PPB in clean limit inevitably becomes the $n=1$ LL vortex lattice (see Fig.6 (a) in Ref.\cite{AI03}). However, the HFSC phase in FeSe in ${\bf H} \parallel c$ case is not the $n=1$ LL state, judging from the experimental fact \cite{Kasa21,Kasa14} that the field-induced structural transition between the conventional vortex solid in lower fields and the HFSC one is continuous. Then, the HFSC phase in ${\bf H} \parallel c$ should be identified with one of the FFLO vortex solids which are described within the $n=0$ LL and hence, have a spatial modulation parallel to the applied field \cite{AI03}. According to Fig.2 in Ref.\cite{AI15}, the $n=1$ LL vortex state may be destabilized even at zero temperature in the cases of {\it two-band} electronic models with an $\alpha_{\rm M}$-value of order unity so that it is possible that the HFSC phase in a two-band system in clean limit is indeed a $n=0$ LL vortex lattice with a spatial modulation parallel to ${\bf H}$. 

In the case of a $n=0$ vortex lattice described by the GL model, the type of the FFLO spatial modulation is found to depend on the nature of the $H_{c2}$-transition at low temperatures \cite{Agter,Nakashima}: When the mean field $H_{c2}$-transition is of second order, the HFSC phase should be a hybrid of the triangular vortex lattice and the phase-modulated FF state \cite{FF} and cannot become the triangular vortex lattice with the LO-like periodic amplitude-modulation \cite{LO} parallel to ${\bf H}$. The triangular vortex lattice with the LO-like spatial modulation parallel to ${\bf H}$ becomes the HFSC phase only when the mean field $H_{c2}$-transition is of first order. Then, one might wonder if the fact \cite{Kasa21} mentioned in sec.I that the $H_{c2}$-transition in FeSe in ${\bf H} \parallel c$ is apparently continuous contradict the observation of a nodal plane perpendicular to ${\bf H}$ at the sample surface \cite{Kasa21} suggesting that the HFSC phase in FeSe in ${\bf H} \parallel c$ should be the LO vortex lattice. However, we find that, once the strong SC fluctuation in FeSe is taken into account, there may be no such contradiction: By using the relations eqs.(\ref{Dyson}), (\ref{selfenergy}), (\ref{n0}), (\ref{bc}), and the data of $V_4$ and $V_6$ for $\alpha_{\rm M}=0.8$ given in Fig.5 of Ref.\cite{AI03}, we obtain Fig.4 expressing the field dependence of the mean-squared SC order parameter $\langle |\Delta|^2 \rangle$ at each temperature, $t=T/T_{c0}=0.2$, $0.15$, $0.1$, $0.05$, and $0.01$. We note that the mean field $H_{c2}$-transition is of first order, i.e., $V_4 < 0$, in $t \leq 0.35$ for the set of the parameters used in Fig.4. Nevertheless, the ${\tilde N}_0$-value of eq.(\ref{n0}) makes the discontinuous change of $\langle |\Delta|^2 \rangle$ in the mean field theory at $H_{c2}(0)$ broad enough in $t \geq 0.05$ (compare Fig.4 with Figs.9 and 11 in Ref.\cite{AI03}). Here, we note that the quantum SC fluctuation has not been taken into account in the present analysis. Inclusion of the quantum fluctuation would broaden even the $\langle |\Delta|^2 \rangle$-curve at $t=0.01$ in Fig.4. By comparing the curves in Fig.4 with the heat capacity data in Ref.\cite{Kasa21}, we conclude that it is difficult to determine the nature of the mean field $H_{c2}$-transition from the experimental data of real FeSe with strong fluctuation in which the $H_{c2}$ is merely a continuous crossover line. Thus, the present theory does not contradict the conclusion in Ref.\cite{Kasa21} identifying the HFSC phase in FeSe in ${\bf H} \parallel c$ as the LO vortex lattice. 

Finally, the magnetic phase diagram of FeSe in ${\bf H} \perp c$ will be briefly discussed. Although the present work was performed by assuming the configuration with a field perpendicular to the basal plane in a Q2D system, the results in sec.IV should be qualitatively applicable even to the ${\bf H} \perp c$ case. First, the $H_m(T)$-curve in Fig.3 is quite similar to the irreversibility line in Ref.\cite{Kasa20} and to the melting line estimated in Ref.\cite{Hardy} in that the $H_m(T)$-curve is convex in the $H$-$T$ phase diagram in spite of showing a broad vortex liquid regime. However, we expect the FFLO phase in the high $H$ and low $T$ corner in Fig.3 to, in this ${\bf H} \perp c$ case, have been replaced by the $n=1$ LL vortex lattice. In Ref.\cite{NNI}, we have shown that a peculiar field dependence of the resistive behavior around $H_{c2}(T)$ at low enough temperatures in ${\bf H} \perp c$ is qualitatively consistent with the resistivity curve resulting from the quantum SC fluctuation not in the familiar $n=0$ LL but in the $n=1$ LL in the case with moderately strong PPB. It strongly suggests that the HFSC phase in ${\bf H} \perp c$ should be the $n=1$ LL vortex solid \cite{Klein,Allan,Matsu}. Then, the vortex liquid controlled by the $n=1$ LL modes of the SC order parameter should be present just above the melting 
line. Theoretical description of such a novel vortex state should be left for a future work. 

\section{Acknowledgement}
We thank Naratip Nunchot for his help on numerical analysis and for reading the original manuscript and Yuji Matsuda and Shigeru Kasahara for discussions on their experimental data. The present work was supported by JSPS KAKENHI (Grant No. JP21K03468).

\section{Appendix}

Our starting model for deriving the Ginzburg-Landau (GL) Hamiltonian is the simplest BCS Hamiltonian with a single electronic band 
\begin{eqnarray}
{\cal H} &=& \sum_{\sigma= \pm 1} \int d^3r \varphi^\dagger_\sigma({\bf r}) \biggl[ \frac{\hbar^2}{2m} \biggl(-{\rm i}\nabla + \frac{\pi}{\phi_0} {\bf A} \biggr)^2 - I \sigma \biggr] \varphi_\sigma({\bf r}) \nonumber \\
&-& |g| \sum_{\bf q} \Psi^\dagger({\bf q}) \Psi({\bf q}), 
\label{he}
\end{eqnarray}
where $\phi_0=\pi \hbar/|e|$ is the flux quantum, $I=\mu_0 H$ is the Zeeman energy, $|g|$ is the attractive interaction strength, and 
\begin{equation}
\Psi({\bf q}) = \frac{1}{2} \sum_{\bf p} \sum_{\sigma = \pm 1} \, \sigma \, w_{\bf p} \, c_{{-{\bf p} + {\bf q}/2}, -\sigma} \, c_{{\bf p}+{\bf q}/2, \sigma} 
\end{equation}
is the pair-field operator expressed by a spin-singlet pairing function $w_{\bf p}$ and $c_{{\bf p},\sigma}$ which is the Fourier transform of the electron operator $\varphi_\sigma({\bf r})$. For simplicity, the $s$-wave paired case with $w_{\bf p}=1$ will be assumed. 
The following GL Hamiltonian is obtained from the electronic model (\ref{he}) through an extension to Q2D case: 
\begin{widetext}
\begin{equation}
{\cal H}_{\rm GL} = N(0) \int d^3{\bf r} \biggl[ \Delta^* (a_0 + B (-\partial_z^2) + C \partial_z^4 ) \Delta + \frac{V_4}{2} |\Delta|^4 + \frac{V_6}{3} |\Delta|^6 \biggr], 
\label{GLAp}
\end{equation}
\end{widetext}
where the order parameter $\Delta$ is assumed to be in the $n=0$ LL subspace, and the applied field is assumed to be perpendicular to the basal plane. The coefficient $a_0$ is given by 
\begin{equation}
a_0= {\rm ln}(t) + \int_0^\infty d\rho \biggl( \frac{2 \pi t}{{\rm sinh}(2 \pi t \rho)} - f(\rho) \exp(-|\nu|^2 \rho^2/2) \biggr), 
\label{a0}
\end{equation}
where $t=T/T_{c0}$, $T_{c0}$ is the zero field SC transition temperature, and 
\begin{equation}
f(\rho)=\frac{2 \pi t}{{\rm sinh}(2 \pi t \rho)} \, {\rm cos}\biggl(2 \frac{I}{T_{c0}} \rho \biggr). 
\end{equation}
The coefficients of the gradient terms are given by $B=b \xi_{0,\parallel}^2$ and $C=c \xi_{0,\parallel}^4$, where 
\begin{eqnarray}
b &=& \int_0^\infty d\rho \rho^2 f(\rho) \exp(-\rho^2|\nu|^2/2), \nonumber \\
c &=& - \frac{1}{4} \int_0^\infty d\rho \rho^4 f(\rho) 
\exp(-\rho^2 |\nu|^2/2), 
\label{bc}
\end{eqnarray}
where $\xi_{0,\parallel}$ is the out-of-plane coherence length at zero 
temperature, and $\nu= \sqrt{2} \pi \xi_0 ({\hat p}_x 
+ {\rm i} {\hat p}_y)/r_H$. 

The limitation to the $n=0$ LL modes imply that the mode coupling terms, the fourth-order and six-order terms in eq.(\ref{GLAp}), are spatially nonlocal. As indicated in Ref.\cite{AI03}, however, this nonlocality seems to be safely negligible for most purposes, and, for simplicity, the local forms of the mode coupling terms were assumed above. In the case of a layered system with a cylindrical Fermi surface, the coefficients $V_4$ and $V_6$ have been derived in Ref.\cite{AI03,RI07} and, in clean limit, are given by 
\begin{widetext}
\begin{eqnarray}
V_4 &=& 3 \, T_{c0}^{-2} \int \Pi_{i=1}^{3} d\rho_i f\biggl(\sum_{j=1}^3 \rho_j \biggr) \biggl\langle \exp\biggl(-\frac{1}{2}\biggl(R_{14}-\frac{1}{2}R_{24} \biggr) \biggr) \, {\rm cos}(I_4) \biggr\rangle_{\rm FS}, \nonumber \\
V_{6} &=& -15 T_{c0}^{-4} \int \Pi_{j=1}^{5} d\rho_j f\biggl(\sum_{j=1}^5 \rho_j \biggr) \biggl\langle \exp\biggl(-\frac{1}{2}(R_{16}+R_{26}) \biggr) \, {\rm cos}(I_6) \biggr\rangle_{\rm FS}, 
\end{eqnarray}
\end{widetext}
where 
\begin{eqnarray}
R_{14} &=& |\nu|^2(\sum_{j=1}{3} \rho_j^2 + \rho_2(\rho_3+\rho_1)), \nonumber \\
R_{24} &=& {\rm Re}(\nu^2) (\rho_2^2 + (\rho_3 - \rho_1)^2) \nonumber \\
I_4 &=& \frac{{\rm Im}(\nu^2)}{4} (\rho_2^2 - (\rho_3 - \rho_1)^2) \nonumber \\
R_{16} &=& |\nu|^2 \biggl(e_1+e_2+e_3 + \frac{2}{3} e_4 e_5 \biggr) 
\nonumber \\
R_{26} &=& {\rm Re}(\nu^2) \biggl(e_1+e_2+e_3 - \frac{e_4^2 + e_5^2}{3} \nonumber \\ 
&-& \frac{2}{3} (e_6 + e_7 + e_8 + e_9) \biggr)  \nonumber \\
I_6 &=& \frac{{\rm Im}(\nu^2)}{4} \biggl(e_1+e_2-e_3 - \frac{e_4^2 - e_5^2}{3} \nonumber \\
&-& \frac{2}{3} (e_6 + e_7 - e_8 - e_9) \biggr)
\end{eqnarray}

\begin{eqnarray}
e_1 &=& (\rho_3+\rho_5)^2 + (\rho_3+\rho_4)^2, \nonumber \\
e_2 &=& (\rho_1+\rho_4+\rho_5)^2, \nonumber \\
e_3 &=& \rho_3^2+\rho_4^2+(\rho_2-\rho_5)^2, \nonumber \\
e_4 &=& \rho_1+2(\rho_3+\rho_4+\rho_5), \nonumber \\
e_5 &=& \rho_2-\rho_3-\rho_4-\rho_5, \nonumber \\
e_6 &=& (\rho_4-\rho_5)^2 + (\rho_1+\rho_5 - \rho_3)^2, \nonumber \\
e_7 &=& (\rho_+\rho_4-\rho_3)^2, \nonumber \\
e_8 &=& (\rho_3-\rho_4)^2 + (\rho_2+\rho_3-\rho_5)^2, \nonumber \\ 
e_9 &=& (\rho_2+\rho_4-\rho_5)^2. 
\end{eqnarray}

To obtain the dispersion relations of the normal modes of the Gaussian fluctuation around $\Delta_0({\bf r})$, the $q$-dependent terms, we follow the Eilenberger's analysis \cite{Eilenberger} to represent the total pair-field in the form $\Delta = e^{{\rm i}q_m z}(\alpha_0 \varphi({\bf r}|0) + a_+ \varphi({\bf r}|{\bf r}_0) e^{{\rm i}qz} + a_- \varphi({\bf r}|-{\bf r}_0) e^{-{\rm i}qz})$. Then, in a situation with $b < 0$, the terms harmonic with respect to $a_\pm$ in ${\cal H}$ take the form 
\begin{widetext}
\begin{eqnarray}
\frac{\delta {\cal H}}{N(0)} &=& [a'_0 + C q^2(q^2+4 q_m^2) + \alpha_0^2(2 {\tilde V}_4 {\overline \xi}_d({\bf r}_0) + 3 {\tilde V}_6 \alpha_0^2 {\overline \eta}_d({\bf r}_0))] (|a_+|^2 + |a_-|^2) + 4C q_m q^3 (|a_+|^2 - |a_-|^2) \nonumber \\
&+& \alpha_0^2[({\tilde V}_4 {\overline \xi}_a({\bf r}_0) + 2 {\tilde V}_6 \alpha_0^2 {\overline \eta}_a({\bf r}_0)) a_+ a_- + {\rm c.c.} ], 
\label{flucfbelow}
\end{eqnarray}
\end{widetext}
where 
\begin{eqnarray}
{\overline \xi}_d({\bf r}_0) &=& \beta_{\rm A}^{-1} \langle |\varphi({\bf r}|0) \varphi({\bf r}|{\bf r}_0)|^2 \rangle_s, \nonumber \\
{\overline \xi}_a({\bf r}_0) &=& \beta_{\rm A}^{-1} \langle (\varphi^*({\bf r}|0))^2 (\varphi({\bf r}|{\bf r}_0))^2 \rangle_s, \nonumber \\
{\overline \eta}_d({\bf r}_0) &=& \gamma_{\rm A}^{-1} \langle |\varphi({\bf r}|0) \varphi({\bf r}|{\bf r}_0)|^2 |\varphi({\bf r}|0)|^2 \rangle_s, \nonumber \\
{\overline \eta}_a({\bf r}_0) &=& \gamma_{\rm A}^{-1} \langle (\varphi^*({\bf r}|0))^2 (\varphi({\bf r}|{\bf r}_0))^2 |\varphi({\bf r}|0)|^2 \rangle_s. 
\label{matrix}
\end{eqnarray}
By performing the "Bogoliubov transformation", the diagonalized form of eq.(\ref{flucfbelow}) becomes $\delta {\cal H} = N(0) (E_+ |{\tilde a}_+|^2 + E_- |{\tilde a}_-|^2)$, where 
\begin{eqnarray}
E_\pm &=& \frac{{\tilde V}_4^2}{4 {\tilde V}_6} \biggl(-s_4+\sqrt{1 - {\overline a}'_0} \biggr) \biggl[ 4 s_4 {\overline \xi}_d 
+ 3\biggl(-s_4 \nonumber \\
&+& \sqrt{1 - {\overline a}'_0}\biggr){\overline \eta}_d \pm 2 \biggl|s_4 {\overline \xi}_a + \biggl(-s_4+\sqrt{1 - {\overline a}'_0} \biggr){\overline \eta}_a \biggr| \biggr] \nonumber \\
&+& a'_0 + C q^4 + 2|B| q^2. 
\label{diagenergy}
\end{eqnarray}
In obtaining eq.(\ref{diagenergy}), O($q^6$) terms were neglected. Here, $\sqrt{2} a_- = \sigma e^{-{\rm i}\gamma/2} {\tilde a}^*_\sigma$, $\sqrt{2} a_+ = e^{-{\rm i}\gamma/2} {\tilde a}_\sigma$ ($\sigma=\pm$), and $\gamma$ is the phase of $s_4 {\overline \xi}_a + (-s_4+\sqrt{1 - {\overline a}'_0}) {\overline \eta}_a$. The corresponding expression in the case with a positive $B$ is given by eq.(\ref{diagenergy}) with the last term replaced by $B q^2$. 

For the massive mode with the excitation energy $E_+$, the ${\bf r}_0$-dependence is not important so that all of the expressions defined in eq.(\ref{matrix}) may be of unity. 
Then, the Hamiltonian expressing the Gaussian amplitude fluctuation 
around the vortex lattice solution may be approximately expressed as 
\begin{eqnarray}
\delta {\cal H}_{\rm amp} &=& N(0) \sum_{q, {\bf k}_\perp} \biggl( \frac{{\tilde V}_4^2}{{\tilde V}_6}\biggl(1 - {\overline a}'_0 - s_4 \sqrt{1 - {\overline a}'_0} \biggr)  \nonumber \\
&+& 2|B|q^2 + C q^4 \biggr)|\delta \Delta_q|^2 
\end{eqnarray}
when $B \leq 0$. This expression will be used in determining the melting line through comparison between the free energy estimated from higher fields and the corresponding one from lower fields. 

As shown previously \cite{Moore,RI591740}, the massless mode with the excitation energy $E_-$ reduces to the purely phase fluctuation in $|{\bf k}_\perp| \to 0$ limit and corresponds to the shear elastic mode of the vortex lattice, where ${\bf k}_\perp = ({\bf r}_0 \times {\hat z})/r_H^2$. 
By examining the ${\bf r}_0$-dependence of the quantities defined in eq.(\ref{matrix}) in details, the Hamiltonian expressing the massless mode is found to become eq.(\ref{phaseenergy}).

\end{document}